\documentclass[12pt]{article}
\usepackage{graphicx}
\usepackage{cite}
\newcommand{\fnd}[2]{\frac{\textstyle #1}{\textstyle #2}}
\newcommand{\xrm}[1]{{\textstyle \mbox{\rm #1}}}
\newcommand{\bmeef}[1]{\mbox{\boldmath $#1$}}

\newcommand{\bra}[1]{\mbox{$\left\langle #1\right|$}}
\newcommand{\ket}[1]{\mbox{$\left| #1\right\rangle$}}
\newcommand{\bracket}[2]{\mbox{$\left\langle #1\left| #2\right.\right
\rangle$}}
\newcommand{\braket}[3]{\mbox{$\left\langle #1\left|
#2\right| #3\right\rangle$}}
\newcommand{\dissum}[2]{\displaystyle \sum_{#1}^{#2}}
\begin{document}
\title{Classification of the scalar mesons: a strange pole expedition into
charm and beauty territory}
\author{
Eef van Beveren\\
{\normalsize\it Centro de F\'{\i}sica Te\'{o}rica}\\
{\normalsize\it Departamento de F\'{\i}sica, Universidade de Coimbra}\\
{\normalsize\it P-3000 Coimbra, Portugal}\\
{\small http://cft.fis.uc.pt/eef}\\ [.3cm]
\and
George Rupp\\
{\normalsize\it Centro de F\'{\i}sica das Interac\c{c}\~{o}es Fundamentais}\\
{\normalsize\it Instituto Superior T\'{e}cnico, Edif\'{\i}cio Ci\^{e}ncia}\\
{\normalsize\it P-1049-001 Lisboa Codex, Portugal}\\
{\small george@ajax.ist.utl.pt}\\ [.3cm]
{\small PACS number(s):
12.40.Yx, 14.40.Aq, 14.40.Lb, 13.25.Es, 13.25.Ft, 13.75.Lb}
}

\maketitle

\begin{abstract}
The classification of scalar and vector mesons is reviewed within the
framework of the Resonance-Spectrum Expansion (RSE). This method allows a
simple and straightforward description of non-exotic meson-meson scattering,
incorporating the effects of quark confinement and OZI-allowed decay in a
fully nonperturbative way. Results for resonances and bound states are
compared to experiment, on the basis of computed pole positions and cross
sections. New predictions for open-charm and -bottom scalar mesons are
presented.

Concretely, observed vector states for $u\bar{u}$-$d\bar{d}$, $s\bar{s}$,
$c\bar{c}$, and $b\bar{b}$ are reproduced, and others are predicted.
In the light scalar sector, the now established two nonets,
one below 1 GeV and one in the region 1.3--1.5 GeV,
are easily described,
through the appearance of extra poles in the $S$ matrix.
The recently found $D_{s0}^{\ast}$(2317) meson is accurately reproduced
by the same mechanism,
as a quasi-bound state in the coupled $c\bar{s}$-$DK$ system.

In $S$-wave $D\pi$ and $B\pi$ scattering, new resonances are foreseen
close to threshold, i.e., a $D_{0}^{\ast}$ at 2.16$\pm$0.05 GeV some 250 MeV
wide, and a $B_{0}^{\ast}$ at 5.47$\pm$0.05 GeV with a width of about 50 MeV.
Additional predictions concern the existence of $b\bar{s}$ and $b\bar{c}$
scalar mesons, stable with respect to OZI-allowed decay to $BK$ and $BD$,
respectively, namely at 5.61$\pm$0.05 GeV resp.\ 6.64$\pm$0.05 GeV.
\end{abstract}

\section{Introduction}
It took mankind only about one century to resolve the mystery
of the spectral lines in visible light reported by Joseph Fraunhofer in 1814
\cite{DKAWMVp193,PTRSL92p365}.
The collection of sufficient data lasted several decades,
during which some progress was made by the discovery of
striking patterns in the spectra.
An important step that provided the key to the analysis of spectra
was the classification of hydrogen lines made by Johann Balmer in 1885
\cite{APC25p80}.
This allowed Niels Bohr \cite{PM26p1} later on to account for those lines,
resulting in a spectacular advance in our understanding of Nature.
Nevertheless, one must recognise that Nature had been very helpful in
providing for a relatively weak interaction between the electromagnetic
field and the massive charged constituents of atoms.
As a consequence, atomic spectra consist of
narrow resonances, which can be associated quite easily
with underlying almost-bound states of electrons and a nucleus.
Moreover, this circumstance also allows a perturbative approach
to photon scattering off atoms.
To lowest order, one may largely ignore the presence of the photon,
and just determine the atom's excitation spectrum of stable bound states.
Transitions can next be determined from the interactions
of these hypothetical bound states with the photon field
\cite{AP79p361}.

A similar strategy breaks down for strong interactions.
For example, spectroscopic structures in meson-meson scattering
are the result of an interplay between
various phenomena which all have comparable intensities,
possibly leading to resonances with large widths.
From the valence-quark picture, in which the flavour quantum numbers
of a meson indicate that
it consists of a quark permanently confined to an antiquark,
one might expect a quasi-bound-state resonance spectrum.
But contrary to atomic structures where extra pairs of electrons
and positrons are energetically very disfavoured,
mesons could easily consist of several pairs of light quarks
and antiquarks as well.
Hence, the permanent creation and annihilation of the latter pairs
must provide an important contribution to the meson dynamics.
Furthermore, contrary to photons, gluons also selfinteract,
which could even give rise to systems just consisting of glue.
At present, however, about half a century after the discovery
of the first baryonic resonances and mesons,
we are still at the stage of classifying particles and resonances
into multiplets of flavour, spin, parity, and $C$-parity.

Most notoriously, the low-lying scalar-meson resonances
are still subject to heated debate.
The isotriplet $a_{0}(980)$ and isosinglet $f_{0}(980)$ resonances,
in $\pi$$\eta$ and $\pi$$\pi$ $S$-wave scattering, respectively,
are experimentally well established,
but their classification in terms of quasi-bound systems of quarks and
antiquarks is still far from agreed upon. Possible interpretations run from
$q\bar{q}$ \cite{PRD26p239,PRD69p014010,PLB454p365,PLB456p80}
and $q^{2}\bar{q}^{2}$ \cite{PRD15p267} states,
to molecules of meson pairs \cite{PRD43p95},
whereas mixtures of such configurations are also considered
\cite{ZPC30p615,ZPC68p647,PLB586p53,HEPPH0310170}.

Purely dynamical pictures for the $a_{0}(980)$ and $f_{0}(980)$ resonances,
solely in terms of mesonic degrees of freedom,
would only make sense if the other scalar flavour configurations,
i.e., the two isodoublets $K^{\ast}_{0}$ and
the remaining isosinglet $f_{0}$, either showed up as standard
Breit-Wigner-type resonances, or could not be extracted from the data at all.
A definite solution for this issue depends, however,
on the definition of \em a correct procedure \em \/for the analysis
of scattering and/or production data.
The actual state of affairs is as follows.

The $\pi\pi$ $S$-wave isoscalar scattering data are in
Ref.~\cite{ZPC30p615}
compared to the results of a model in which confined nonstrange (up or down)
and strange quark-antiquark pairs are coupled
to various two-meson channels containing pseudoscalar and vector mesons.
The model has no input $\sigma$-resonance parameters.
Nevertheless, the amplitude exhibits a clear pole at
$\sqrt{s}=(470-208i)$ MeV.
N.~A.~T\"{o}rnqvist and M.~Roos come to a similar conclusion from
a model which does not contain vector-meson pairs
\cite{PRL76p1575}.
So far, many works have confirmed a
scattering pole in $\pi\pi$ $S$-wave isoscalar scattering
\cite{KEK2003-7p109,KEK2003-10p11,PRD69p074008,NPA733p235,
PR389p61,PRD69p074033,HEPPH0310312,HEPPH0310186,HEPPH0309228,
AIPCP670p59,HEPPH0310062,PRD69p034005,KEK2003-7p143,KEK2003-7p186,
KEK2003-7p20,NPA724p357,AIPCP688p45,AIPCP687p74,NPA727p353,JPG30p663,
PRD68p013008,HEPPH0304075,HEPPH0302133,PRD67p054001,HEPPH0304177,
EPJC26p253,DeirdreMargaretBlack,EPJA16p229,JPG28pR249,PLB572p1}.

When the model of Ref.~\cite{ZPC30p615} is applied to
$K\pi$ $S$-wave isodoublet scattering,
the resulting amplitude shows a pole at
$\sqrt{s}=(727-263i)$ MeV, which can obviously be associated with the
$K^{\ast}_{0}(800)$ (or $\kappa$)
resonance found in experiment by the E791 collaboration \cite{PRL89p121801}.
However, the amplitude of T\"{o}rnqvist and Roos \cite{PRL76p1575}
does not show a pole in that region,
whereas S.~N.~Cherry and M.~R.~Pennington \cite{NPA688p823} even
claimed that there is no ``$\kappa(900)$'' resonance.
Nevertheless, many works confirm a scattering pole in
the amplitude for $K\pi$ $S$-wave isodoublet scattering
\cite{HEPPH0406112,KEK2003-7p135,PLB572p1,KEK2003-10p29,JHEP0402p047,
PRL92p102001,HEPPH0302062,PRD67p034025,EPJC22p493,PTP102pE52,NPB587p331,
KEK2003-7p143,1IWFSp0,HEPEX0012009,PRD60p074023,PRD58p054012,PRD59p074001,
PLB413p137,HEPPH9712230,PTP98p621,PTP95p745,PRD26p239,NPB60p233,
PR184p1609,PRD4p1018,PLB28p203,PR168p1708}.

For other flavours one expects similar $S$-wave resonances.
Neveretheless, since charm and bottom are much heavier than the light flavours,
the resulting signals may be very differently situated with respect to
the dominant thresholds.
The $f_{0}(600)$ (or $\sigma$) and $K^{\ast}_{0}(800)$ structures
lie well above the $\pi\pi$ and $K\pi$ thresholds, respectively,
whereas the $a_{0}(980)$ and $f_{0}(980)$ resonances are very close
to the $KK$ threshold.
Their (preliminary) charmed-nonstrange partner $D$(2290) shows up as a
resonance in $S$-wave $D\pi$ \cite{HEPEX0307021,HEPEX0210037}
some 200 MeV above threshold.
But their recently oberved charmed-strange partner $D^{\ast}_{s0}(2317)$
\cite{PRL90p242001,HEPEX0305100,HEPEX0308019} lies about
45 MeV below the $DK$ threshold.
Hence, when the very light pions are involved,
the lowest-lying $S$-wave signals come out well above threshold,
while for other flavours the signals turn up as bound states
with respect to OZI-allowed decays.
We shall study this behaviour in more detail next.

\section{Is there no \bmeef{K^{\ast}_{0}(800)}?}

In 1996, T\"{o}rnqvist and Roos \cite{PRL76p1575}
wrote: ``{\it ... one can understand the data on the lightest scalars
in a model which includes most well-established theoretical constraints:
{\bf Adler zeroes} as required by chiral symmetry,
all light two-pseudoscalar thresholds with {\bf flavour symmetric couplings},
physically acceptable analyticity, and unitarity.
A unique feature of the model is that it simultaneously describes
{\bf the whole scalar nonet},
and one obtains a good representation of a large set of relevant data.
Only {\bf six} parameters, which all have a clear physical interpretation,
were needed}.''
In their work they discussed a phenomenon called {\bf pole doubling}, which
reveals signals in the scattering amplitude that do not
directly stem from an underlying quark-antiquark spectrum for scalars.
T\"{o}rnqvist and Roos found that the $f_{0}(600)$, or $\sigma$,
as well as the $a_{0}(980)$ and isosinglet $f_{0}(980)$ resonances
exactly represent such phenomena.
But with respect to the isodoublets they wrote: ``{\it ... in the strange
channel there is only one important channel open, the $K\pi$ with
a Clebsch-Gordan coefficient {\bf reducing the coupling} compared
to $s\bar{s}$-$K\bar{K}$ by $\sqrt{3/4}$.
{\bf This}, together with the fact that the $K\pi$ threshold involves
two unequal mass mesons, implies that the resonance doubling phenomenon
{\bf does not appear in the strange sector}.}''

\begin{figure}[hbpt]
\centerline{\scalebox{0.7}{\includegraphics{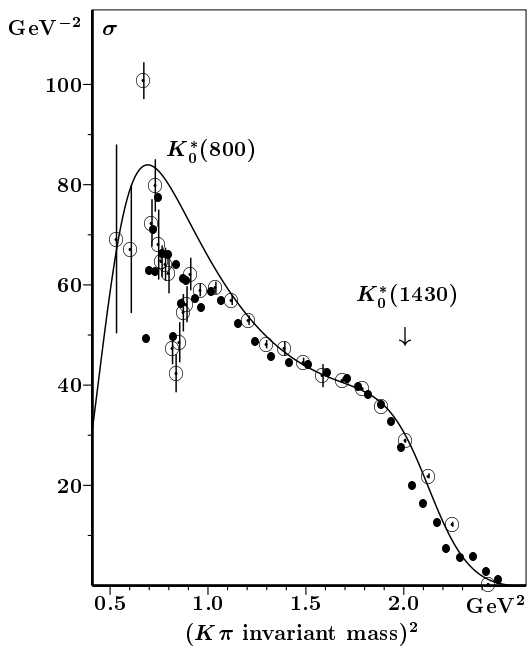}}}
\caption[]{The cross section for elastic $K\pi$ scattering in $S$ waves
from model \cite{EPJC22p493}, compared to
experiments \cite{NPB133p490,PRD19p2678} ($\odot$) and
\cite{NPB296p493} ($\bullet$).}
\label{KpiS}
\end{figure}

Nevertheless, in Ref.~\cite{EPJC10p469} we pointed out that flavour invariance
does {\bf not} lead to the coupling constants determined by
T\"{o}rnqvist in Ref.~\cite {ZPC68p647}.
As a consequence, the $K\pi$ $S$-wave scattering amplitude develops
a scattering pole which is highly non-Breit-Wigner, showing up
as a signal in the cross-section (see Fig.~\ref{KpiS}) for the unitarised
model published already in 1986 \cite{ZPC30p615}.
This model has all the features
proclaimed by T\"{o}rnqvist and Roos, except for the phenomenological Adler
zeroes.
However, it is easy to show that the employed relativistics kinematics for the
two-meson channels leads to effective Adler zeroes below threshold, with the
right properties, without any need to introduce such {\em ad hoc} \/factors
by hand \cite{HEPPH0310320,BBKR04}.
Moreover, also vector-vector thresholds were included in Ref.~\cite{ZPC30p615}.

In 2001, Cherry and Pennington \cite{NPA688p823}
wrote: ``{\it By using a model-independent analytic continuation of
the existing data on $\pi^{+}K^{-}$ scattering between 850 MeV
and 2.5 GeV, we find that there is only one scalar resonance below
1600 MeV, which is readily identified with the $K^{\ast}_{0}(1430)$.
Our procedure requires no assumptions about the amplitude being described
by Breit-Wigner forms with any particular form of background.
It directly counts the number of poles of the $S$-matrix on the nearby
unphysical sheet.
This method provides a rigorous test of whether the {\boldmath $\kappa(900)$}
exists, and we conclude {\bf it does not}.
This is so whether we use the older Estabrooks et al.}
\cite{NPB133p490} {\it data
or that of the LASS colaboration} \cite{NPB296p493}.''

However, in 2002, M.~Boglione and M.~R.~Pennington \cite{PRD65p114010} found
structures in the $K\pi$ $S$-wave scattering amplitude below 1.0 GeV, but still
concluded that ``{\it a light $\kappa$ scalar meson cannot be generated as a
conventional resonance but only as a bound state}''. Nevertheless, they now
also concluded that, for $S$-wave scattering in general, one can generate more
than one physical resonance starting from only one bare $q\bar{q}$ state.

In 2004, C.~Amsler and N.~A.~T\"{o}rnqvist \cite{PR389p61}
wrote: ``{\it The nonlinear form of $\Pi (s)$ can produce two poles
in the amplitude, although only one seed state
($q\bar{q}$ or 4-quark state) is introduced.
\ldots there are theoretical arguments for why a light and broad
pole can exist near the $K\pi$ threshold and many phenomenological papers
support its existence.
But the question of whether a $K^{\ast}_{0}(800)$ exists near the $K\pi$
threshold is not yet conclusive, since Breit­Wigner fits have been used.
We believe that experimental groups should look for pole positions
in their data analysis, which also include the aforementioned nonlinear
effects from $S$-wave thresholds.}''

From the foregoing, it seems prudent to conclude that a $K_0^*$-resonance pole
below 1 GeV may very well exist, be it a $\kappa(800)$ or a $\kappa(900)$,
which is all we need in order to unambiguously classify all scalar mesons, as
we shall show below.

\section{The Resonance-Spectrum Expansion}

The forces supposed to hold together quarks and antiquarks
inside a hadron are, in the majority of models, described by a potential
having the observed properties of strong interactions,
like confinement and flavour independence.
All we need here is to assume the existence of some confinement mechanism
which has associated with it a complete set of permanently bound states.
In the interaction region for non-exotic meson-meson scattering,
the contribution of OZI-allowed quark-pair creation and annihilation
dominates the hadronic processes,
such that temporarily the system consists of only
one valence $q\bar{q}$ pair.
This intermediate state can be written as an expansion in terms of the
confinement bound states.

Let us denote by $H_{c}$ the operator which describes the dynamics of
confinement and, furthermore,
its properly normalised eigensolutions,
corresponding to the energy eigenvalue $E_{n\ell_{c}}$, by

\begin{equation}
\bracket{\vec{r}\,}{n\ell_{c}m_{c}}\; =\;
Y_{\ell_{c}m_{c}}\left(\hat{r}\right)\;
{\cal F}_{n\ell_{c}}(r)
\;\;\; ,\;\;\;
\xrm{with}\;\;\;
\left\{
\begin{array}{l}
\xrm{$n=0$, $1$, $2$, $\dots$}\\
\xrm{$\ell_{c}=0$, $1$, $2$, $\dots$}\\
\xrm{$m_{c}=-\ell_{c}$, $\dots$, $+\ell_{c}$}
\end{array}
\right.
\;\;\;\; .
\label{Vccomplete}
\end{equation}

\noindent
Note that we assume radial symmetry for confinement.

Far from the interaction region, the system consists of a pair of free
mesons, the dynamics of which we denote by $H_{f}$.
Elastic scattering of meson pairs is described by the transition operator

\begin{equation}
T\; =\;\left(\; 1\; -\; VG_{f}\;\right)^{-1}\; V
\; =\;
\sum_{N=0}^{\infty}\;\left( VG_{f}\right)^{N}\; V
\;\;\; ,
\label{Titeration}
\end{equation}

\noindent
where $G_{f}$ represents the free boson propagator
associated with the operator $H_{f}$ and given by

\begin{equation}
G_{f}\left({{\vec{k}\,}'\;},{\vec{k}\;}; z\right)\; =\;
\braket{{{\vec{k}\,}'\;}}{\left( z-H_{f}\right)^{-1}}{\vec{k}\;}\; =\;
\fnd{2\mu}{2\mu z-k^{2}}\;
\delta^{(3)}\left(\vec{k}\, -{\vec{k}\,}'\right)
\;\;\; .
\label{Greensfu}
\end{equation}

The various mass parameters employed in this section are listed in
Table~\ref{masses}.

\begin{table}[hbpt]
\begin{center}
\begin{tabular}{|c||l|}
\hline\hline
symbol &
definition \\
\hline
$m_{q}$ $\left( m_{\bar{q}}\right)$ & constituent quark (antiquark) mass \\
$\mu_{c}$ & reduced mass of $q\bar{q}$ system \\
$M_{1,2}$ & meson masses \\
$\mu$ & reduced two-meson mass \\
\hline\hline
\end{tabular}
\end{center}
\caption[]{Mass parameters used in this section.}
\label{masses}
\end{table}

If we neglect possible two-meson final-state interactions,
we may represent the communication between the intermediate $q\bar{q}$ state
and the free two-meson states by an interaction potential $V_{t}$,
which essentially describes three-meson vertices.
The operator $V_{t}$ mediates between systems
with different quantum numbers,
as from parity conservation we deduce that the relative orbital angular
momenta $\ell$ of a meson-meson pair and $\ell_{c}$ of the confined
quark-antiquark pair differ by at least one unit,
whereas, moreover,
the total intrinsic spins are different for the two systems when
$C$-parity is taken into account.
The conservation of $J$, $P$, $C$, isospin, flavour, and colour can be
formulated in a consistent way \cite{ZPC21p291,ZPC17p135,HEPPH0305056},
but we do not intend to go into details here.
We just recall that there exists relations between
($\ell_{c}$, $m_{c}$) for the various intermediate $q\bar{q}$ states,
and ($\ell$, $m$) for the various two-meson states,
and, furthermore, that the convolution integrals expressing such relations
result in relative coupling constants for each of the possible
three-mesons vertices \cite{EPJC11p717}.
The radial parts of the transition potentials can also be deduced
in this formalism \cite{ZPC21p291}.
Resuming, we write
\begin{equation}
V_{t}\; =\;\lambda\; U\left( r\right)
\;\;\; ,
\label{Vt}
\end{equation}
where $\lambda$ stands, in general, for a matrix containing the relative
three-meson coupling constants.
In the one-channel case, it parametrises the coupling constant of the
three-meson vertex involved.

For the Born term of expression (\ref{Titeration}), we then obtain
\vspace{-25pt}

\begin{equation}
\bra{\vec{p}\,}\; V\;\ket{{\vec{p}\,}'}\; =\;
\begin{picture}(44,40)(0,0)
\includegraphics{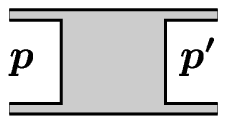}%
\end{picture}\; =\;
\bra{\vec{p}\,}\; V_{t}\;
\left( E(p)-H_{c}\right)^{-1}\; V_{t}\;
\ket{{\vec{p}\,}'}
\;\;\; ,
\label{MSgeneralpot}
\end{equation}
where the total center-of-mass energy $E$
and the linear momentum $p$ are related by
\begin{equation}
E(p)\; =\;
\fnd{{\vec{p}\;}^{2}}{2\mu}\; +\; M_{1}\; +\; M_{2}
\;\;\; .
\label{Ep}
\end{equation}

The shaded area in the diagram of formula (\ref{MSgeneralpot})
is to remind us that quarks and antiquarks are confined,
which mechanism produces the $s$-channel resonances in nature.
In our model, we may study the contribution to those resonances
of any of the confinement states, not just one of them.
It leads us to expressions for the scattering amplitudes
which substantially differ from Breit-Wigner sums.
Our diagrams do not rigorously represent the mathematics involved,
but are merely meant to graphically interpolate
between our understanding of the physical processes and our model.
Actually, instead of shading the areas between the quark and antiquark
lines for the initial- and final-state mesons, we should have represented
each of them by just one line in the diagram of formula (\ref{MSgeneralpot}),
since they are considered point particles in our model.

By letting the self-adjoint operator $H_{c}$ act to the left in
Eq.~(\ref{MSgeneralpot}), we find
\begin{equation}
\bra{\vec{p}\,}\; V\;\ket{{\vec{p}\,}'}\; =\;
\sum_{n\ell_{c}m_{c}}\;\bra{\vec{p}\,}\; V_{t}\;
\fnd{\ket{n\ell_{c}m_{c}}\;\bra{n\ell_{c}m_{c}}}{E(p)-E_{n\ell_{c}}}
\; V_{t}\;\ket{{\vec{p}\,}'}
\;\;\; .
\label{MSgeneralpot1}
\end{equation}
The individual matrix elements in the sum of formula~(\ref{MSgeneralpot1})
result in
\begin{equation}
\bra{n\ell_{c}m_{c}}\; V_{t}\;\ket{\vec{k}\,}\; =\;
\lambda\; (i)^{\ell}\;{\cal J}_{n\ell}(k)\;
Y^{\ast}_{\ell m}\left(\hat{k}\right)
\;\;\; ,
\label{nlmVk}
\end{equation}
where, in order to simplify the expressions, we have defined the quantities
\begin{eqnarray}
{\cal J}_{n\ell}(p) & = & \sqrt{\fnd{2}{\pi}}\;
\int r^{2}dr\; j_{\ell}\left( pr\right)\; U\left( r\right)\;
{\cal F}^{\ast}_{n\ell_{c}}\left( r\right)
\;\;\; ,
\nonumber \\[2mm]
{\cal N}_{n\ell}(p) & = & \sqrt{\fnd{2}{\pi}}\;
\int r^{2}dr\; n_{\ell}\left( pr\right)\; U\left( r\right)\;
{\cal F}^{\ast}_{n\ell_{c}}\left( r\right)
\;\;\;,\;\;\;\xrm{and}
\label{calJcalH}
\\ [2mm]
{\cal H}^{(1,2)}_{n\ell}(p) & = &
{\cal J}_{n\ell}(p)\;\pm i\; {\cal N}_{n\ell}(p)\; =\;
\sqrt{\fnd{2}{\pi}}\;
\int r^{2}dr\; h^{(1,2)}_{\ell}\left( pr\right)\; U\left( r\right)\;
{\cal F}^{\ast}_{n\ell_{c}}\left( r\right)
\;\;\; .
\nonumber
\end{eqnarray}

For the full Born term (\ref{MSgeneralpot1}), we obtain
\begin{eqnarray}
\bra{\vec{p}\,}\; V\;\ket{{\vec{p}\,}'} & = &
\fnd{\lambda^{2}}{4\pi}\;\sum_{\ell}\; (2\ell +1)
P_{\ell}\left(\hat{p}\cdot{\hat{p}\,}'\right)\;
\sum_{n}\;
\fnd{{\cal J}^{\ast}_{n\ell}(p)\; {\cal J}_{n\ell}(p')}{E(p)-E_{n\ell_{c}}}
\;\;\; .
\label{MSgeneralpot2}
\end{eqnarray}

Similarly, one may determine the $N\!+\!1$-st-order term
$\bra{\vec{p}\,}\;\left( VG_{f}\right)^{N+1}\;\ket{{\vec{k}\,}}$
of Eq.~(\ref{Titeration}), yielding
\vspace{-25pt}

\begin{displaymath}
\bra{\vec{p}\,}\;\left( VG_{f}\right)^{N+1}\;\ket{{\vec{p}\,}'}\; =\;
\begin{picture}(212,40)(0,0)
\includegraphics{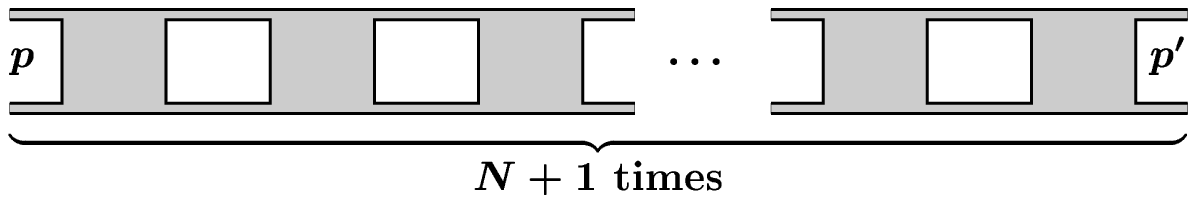}%
\end{picture}
\end{displaymath}
\vspace{10pt}

\begin{eqnarray}
& = &
\fnd{\lambda^{2}}{4\pi}\;
\sum_{\ell}\; (2\ell +1)\;
\left[
\left( -i\;\pi\lambda^{2}\mu p\right)\;
\sum_{n}\;
\fnd{{\cal J}^{\ast}_{n\ell}(p)\; {\cal H}^{(1)}_{n\ell}(p)}
{E(p)-E_{n\ell_{c}}}
\right]^{N}\;
\times\nonumber\\ [.3cm] & & \;\;\;\;\;\;\;\;\times\;
P_{\ell}\left(\hat{p}\cdot{\hat{p}\,}'\right)\;
\sum_{n'}\;
\fnd{{\cal J}^{\ast}_{n'\ell}(p)\; {\cal J}_{n'\ell}(p')}
{E(p)-E_{n'\ell_{c}}}\;
\fnd{2\mu}{2\mu_{f} z-{p'}^{2}}\;
\;\;\; .
\label{termNp1}
\end{eqnarray}
From Eqs.~(\ref{MSgeneralpot2}) and (\ref{termNp1}), we find
for the matrix elements of the product $\left( VG_{f}\right)^{N}\; V$
the expression
\begin{eqnarray}
\lefteqn{\int d^{3}k\;
\bra{\vec{p}\,}\;\left( VG_{f}\right)^{N}\;\ket{{\vec{k}\,}}\;
\bra{\vec{k}\,}\; V\;\ket{{\vec{p}\,}'}\; =}
\nonumber\\ [.3cm] & &
\fnd{\lambda^{2}}{4\pi}\;
\sum_{\ell}\; (2\ell +1)\;
\left[
\left( -i\;\pi\lambda^{2}\mu p\right)\;
\sum_{n}\;
\fnd{{\cal J}^{\ast}_{n\ell}(p)\; {\cal H}^{(1)}_{n\ell}(p)}
{E(p)-E_{n\ell_{c}}}
\right]^{N}\;
\times\nonumber\\ [.3cm] & & \;\;\;\;\times\;
P_{\ell}\left(\hat{p}\cdot{\hat{p}\,}'\right)\;
\sum_{n'}\;
\fnd{{\cal J}^{\ast}_{n'\ell}(p)\; {\cal J}_{n'\ell}(p')}
{E(p)-E_{n'\ell_{c}}}\;
\;\;\; .
\label{TN}
\end{eqnarray}
The terms labelled with $N$ in Eq.~(\ref{TN}) can be summed up, resulting in
a compact final formula for the matrix elements of the transition
operator $T$ in Eq.~(\ref{Titeration}), reading
\vspace{-25pt}

\begin{displaymath}
\bra{\vec{p}\,} T\ket{{\vec{p}\,}'}\; =\;
\begin{picture}(275,40)(0,0)
\includegraphics{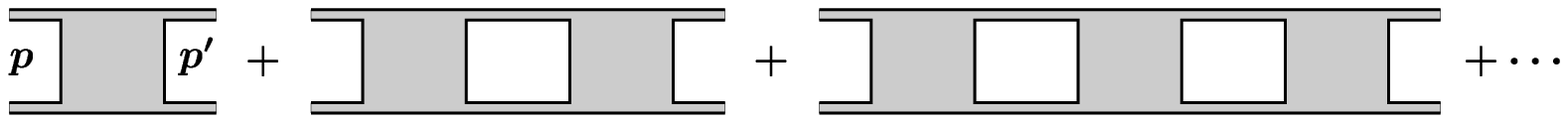}%
\end{picture}
\end{displaymath}
\vspace{10pt}

\begin{equation}
\;\;\; =\;
\fnd{\lambda^{2}}{4\pi}\;
\sum_{\ell =0}^{\infty}(2\ell +1)\;
P_{\ell}\left(\hat{p}\cdot{\hat{p}\,}'\right)\;
\fnd{\dissum{n=0}{\infty}
\fnd{{\cal J}^{\ast}_{n\ell}(p){\cal J}_{n\ell}(p')}
{E(p)-E_{n\ell_{c}}}}
{1+i\pi\lambda^{2}\mu p
\left(\fnd{\lambda a}{\mu_{c}}\right)^{2}
\dissum{n=0}{\infty}
\fnd{{\cal J}^{\ast}_{n\ell}(p){\cal H}^{(1)}_{n\ell}(p)}
{E(p)-E_{n\ell_{c}}}} \; .
\label{T}
\end{equation}

In the limit of very small coupling constant $\lambda$, we recover
from formula (\ref{T}) the usual sum of Breit-Wigner resonances.
But for couplings representing strong interactions,
expression (\ref{T}) behaves very differently.
In the first place, one obtains resonances which are substantially
shifted away from the confinement spectrum $E_{n\ell_{c}}$,
demonstrating the difficulty to extract a confinement spectrum
directly from the data.
Then, one finds poles in the amplitudes which do not stem
from the confinement spectrum,
but find their origin in the scattering continuum.
The structures in meson-meson scattering associated with these poles
are unexpected, which certainly does not facilitate their classification.
A final property of the amplitude (\ref{T}) is related to
its analytic continuation below threshold.
There, one finds bound-state poles on the real-energy axis
associated with mesons which are stable with respect to OZI-allowed
strong decay.
Consequently, all mesons, resonances and bound states, can be generated
with one analytic expression \cite{HEPPH0201006}.

\section{Vector mesons}

The two parameters $\lambda$ and $a$ of the transition potential $V_{t}$
can be fixed, together with the constituent charm and bottom masses,
by the spectra of the $c\bar{c}$ and $b\bar{b}$ vector states.
A comparison of model results \cite{PRD21p772,PRD27p1527}
with experiment is shown in Table~\ref{ccbb}.
\begin{table}[htbp]
\begin{center}
\begin{tabular}{|c||rrr|rrr|}
\hline\hline & & & & & & \\ [-8pt]
& \multicolumn{3}{c|}{$c\bar{c}$}
& \multicolumn{3}{c|}{$b\bar{b}$} \\
state & model & exp. & Ref. & model & exp. & Ref.\\
\hline & & & & & & \\ [-8pt]
& GeV & GeV & & GeV & GeV & \\ [3pt]
1$S$ & 3.10 & 3.10 & \cite{PRD66p010001}
& 9.41 & 9.46 & \cite{PRD66p010001}\\
2$S$ & 3.67 & 3.69 & \cite{PRD66p010001}
& 10.00 & 10.02 & \cite{PRD66p010001}\\
1$D$ &  3.80 & 3.77 & \cite{PRD66p010001}
& 10.14 & 10.16 & \cite{HEPEX0404021}\\
3$S$ & 4.05 & 4.04 & \cite{PRD66p010001}
& 10.40 & 10.36 & \cite{PRD66p010001}\\
2$D$ &  4.14 & 4.16 & \cite{PRD66p010001}
& 10.48 & ...  & \\
4$S$ &  4.41 & 4.42 & \cite{PRD66p010001}
& 10.77 & 10.58 & \cite{PRD66p010001}\\
3$D$ &  ...  & ... & & 10.86 & 10.87 & \cite{PRD66p010001}\\
5$S$ &  ...  & ... & & 11.15 & 11.02 & \cite{PRD66p010001}\\
\hline\hline
\end{tabular}
\end{center}
\caption[]{Comparison of the model results \cite{PRD21p772,PRD27p1527}, for
$J^{PC}\!=\!1^{--}$ charmonium and bottomonium bound-state masses and
the real parts of resonance pole positions, to experiment.}
\label{ccbb}
\end{table}
For confinement we employ here a flavour-independent harmonic-oscillator
force with frequency $\omega\approx 0.19$ GeV.
Consequently, the confinement spectrum has the same level splitting
for all flavours.
If $\nu +1$ represents the radial, and $\ell_{c}$ the angular excitation,
then the total bare quarkonium mass $E_{\nu\ell_{c}}$ is given by
\begin{equation}
E_{\nu\ell_{c}}\; =\;\omega\left( 2\nu +\ell_{c} +\frac{3}{2}\right)\;
+\;\xrm{quark masses}
\;\;\; .
\label{Eellnu}
\end{equation}
In Table~\ref{qqbarMM} we show the meson-meson channels
which are involved in the determination of the quarkonium spectra.
Vector-vector channels come in pairs, one for total spin zero
and one for total spin equal to two. Moreover, spin 2 may combine with $F$
waves.
\begin{table}[hbpt]
\begin{center}
\begin{tabular}{|c||c|c|c|c||}
\hline\hline & & \multicolumn{3}{c|}{ } \\ [-8pt]
& $q\bar{q}$ & \multicolumn{3}{c||}{meson-meson}\\ [3pt]
(L,S) & (0,1) (2,1) & (1,0) & (1,1) & (1,0) (1,2) (3,2)\\ [3pt]
\hline\hline & & & & \\ [-8pt]
$\rho$ &
$n\bar{n}$ &
$\pi\pi$, $KK$ &
$\pi\omega$, $\eta\rho$, $KK^{\ast}$ &
$\rho\rho$, $K^{\ast}K^{\ast}$\\
\hline & & & & \\ [-8pt]
$K^{\ast}$ &
$s\bar{n}$ &
$\pi K$, $\eta K$, $\eta 'K$ &
$\pi K^{\ast}$, $\eta K^{\ast}$, $\eta 'K^{\ast}$ &
$\rho K^{\ast}$, $\omega K^{\ast}$, $\phi K^{\ast}$\\
& & & $K\rho$, $K\omega$, $K\phi$ & \\
\hline & & & & \\ [-8pt]
$\omega$, $\phi$ &
$n\bar{n}$, $s\bar{s}$ &
$\eta\omega$, $KK$ &
$\pi\rho$, $\eta '\phi$, $KK^{\ast}$ &
$K^{\ast}K^{\ast}$ \\
\hline & & & & \\ [-8pt]
$\psi$ &
$c\bar{c}$ &
$DD$, $D_{s}D_{s}$  &
$DD^{\ast}$, $D_{s}D_{s}^{\ast}$  &
$D^{\ast}D^{\ast}$, $D_{s}^{\ast}D_{s}^{\ast}$ \\
\hline & & & & \\ [-8pt]
$\Upsilon$ &
$b\bar{b}$ &
$BB$, $B_{s}B_{s}$ &
$BB^{\ast}$, $B_{s}B_{s}^{\ast}$ &
$B^{\ast}B^{\ast}$, $B_{s}^{\ast}B_{s}^{\ast}$ \\
\hline\hline
\end{tabular}
\end{center}
\caption[]{The channels involved in the determination of
the pole positions.}
\label{qqbarMM}
\end{table}
The complete wave function of charmonium (bottomonium) is thus composed of
twelve channels, the two $c\bar{c}$ ($b\bar{b}$) channels,
one for $S$ wave and one for $D$ wave, coupled to ten meson-meson channels
through $^{3}P_{0}$ transitions. Note that not all of these channels are open
for the systems under consideration, but this does not necessarily mean their
influence is negligible.
The relative coupling intensities for the various channels,
which follow from the three-meson vertices defined in
Refs. \cite{ZPC17p135,ZPC21p291}, are tabulated in Ref.~\cite{PLB454p165}.
In Nature,
many more meson-meson channels couple, in principle, but in the present
version of the model only pseudoscalar and vector mesons are considered
in the initial and final scattering states, since these are generally the ones
that lie close enough to have an appreciable effect.

The non-strange and strange constituent quark masses are
fixed by the elastic $P$-wave $\pi\pi$ and $K\pi$ scattering phase shifts,
respectively.
In Table~\ref{mass} we summarise the model results \cite{PRD27p1527}
for the light vector-meson sector.
\begin{table}[htbp]
\begin{center}
\begin{tabular}{||c||r|rr||r|rr||r|rlr||}
\hline\hline
& \multicolumn{3}{c||}{}
& \multicolumn{3}{c||}{}
& \multicolumn{4}{c||}{}\\ [-8pt]
& \multicolumn{2}{c}{$I=1$} &
& \multicolumn{2}{c}{$I=\frac{1}{2}$} &
& \multicolumn{2}{c}{$I=0$} & & \\ [3pt]
state & model & \multicolumn{2}{c||}{experiment}
& model & \multicolumn{2}{c||}{experiment}
& model & \multicolumn{3}{c||}{experiment}\\
\hline & & & & & & & & & & \\ [-8pt]
& GeV & GeV & Ref. & GeV & GeV & Ref. & GeV & GeV & & Ref. \\ [3pt]
1$S$ & 0.76 & 0.77 & \cite{PRD66p010001}
& 0.93 & 0.89 & \cite{PRD66p010001}
& 0.84 & 0.78 & $\omega(783)$ & \cite{PRD66p010001}\\
& & & & & &
& 1.03 & 1.02 & $\phi(1020)$ & \cite{PRD66p010001}\\ [3pt]
2$S$ & 1.29 & 1.29 &
\cite{PLB408p476}
& 1.41 & 1.41 & \cite{PRD66p010001}
& 1.29 & 1.20 & $\omega(1200)$ & \cite{PLB462p365}\\
& & & & & & & 1.53 & & $\phi$ & \\ [3pt]
1$D$ & 1.40 & 1.42 & \cite{JINRE288521}
& & &
& 1.40 & 1.42 & $\omega(1420)$ & \cite{PRD66p010001}\\
& & & & & & & 1.64 & 1.68 & $\phi(1680)$ & \cite{PRD66p010001}\\ [3pt]
3$S$ & 1.59 & 1.60 & \cite{SLACPUB5606}
& 1.73 & 1.72 & \cite{PRD66p010001}
& 1.67 & 1.65 & $\omega(1650)$ & \cite{PRD66p010001}\\
& & & & & & & 1.87 & & $\phi$ & \\ [3pt]
2$D$ & 1.68 & 1.68 & \cite{JINRE288521} & & & & & & & \\
\hline\hline
\end{tabular}
\end{center}
\caption[]{Real parts of the singularities in the meson-meson scattering
matrices for some of the well-known light vector resonances.}
\label{mass}
\end{table}
We find good agreement between model and experiment.
However, with respect to the radial excitations of the $\rho (770)$ meson,
we believe that the particles refered to in Ref.~\cite{PRD66p010001}
are the $D$ states.
In the harmonic-oscillator spectrum, a $D$ state is degenerate with
the $S$ state belonging to the next radial excitation.
Through the coupling to the meson-meson channels, the degeneracy is lifted,
which gives rise to separate resonances.
Also the $\phi (1680)$ corresponds, in our model, to the $\phi (1D)$ state.

It is remarkable that a model which describes the $c\bar{c}$ and
$b\bar{b}$ quarkonia bound states and resonances very well can predict, to a
fair precision, the $\pi\pi$ and $K\pi$ scattering data, just by fixing
reasonable values for $m_{n}$ and $m_{s}$.

\section{The scalar mesons}

Expression (\ref{T}) can be used to determine
the theoretical scattering phase shifts and cross sections
for elastic and inelastic non-exotic meson-meson scattering.
Results similar to the $K\pi$ elastic cross section shown in
Fig.~\ref{KpiS} have been published in
Refs.~\cite{PRD27p1527,ZPC30p615,EPJC22p493}.
Although it demonstrates that the model describes experiment well,
it does not directly solve important questions on the classification
of, in particular, the light scalar mesons,
because structures which are contested to be genuine resonances in the
experimental cross sections and/or phase shifts, do come out as such in the
model. However, due to the simplicity of the model, which allows
solutions in closed analytic form (\ref{T}),
we can also easily study the pole structure of
the resulting scattering amplitudes.
This appears to be the clue towards
classifying the light scalar mesons,
since it answers the question as to whether there exist isodoublet
and additional isoscalar phenomena related to the $a_{0}(980)$
isotriplet and $f_{0}(980)$ isosinglet resonances.

\begin{figure}[htbp]
\begin{center}
\begin{tabular}{c}
\centerline{\scalebox{0.6}{\includegraphics
{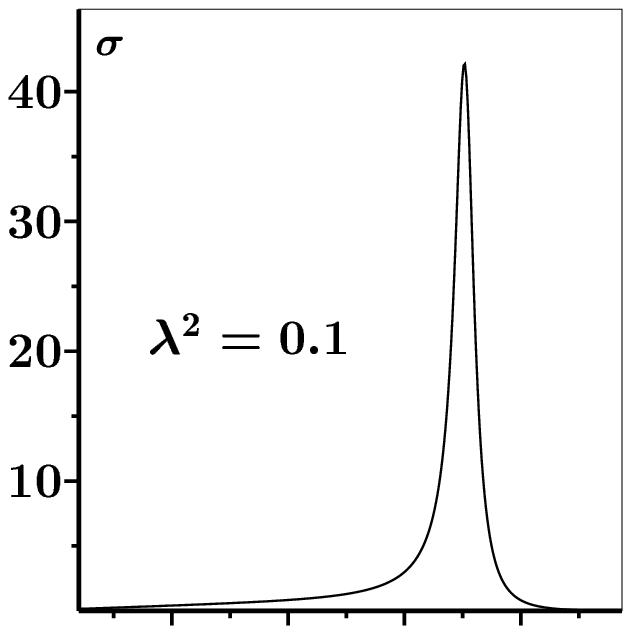}}
\scalebox{0.6}{\includegraphics
{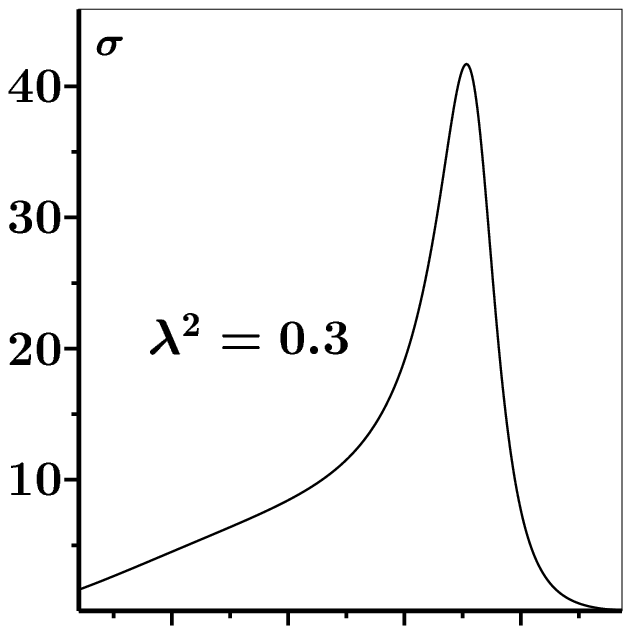}}
\scalebox{0.6}{\includegraphics
{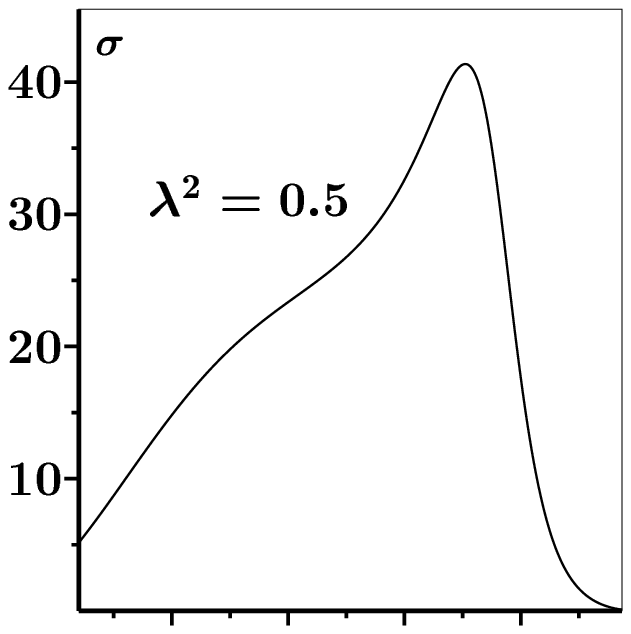}}}\\
\centerline{\scalebox{0.6}{\includegraphics
{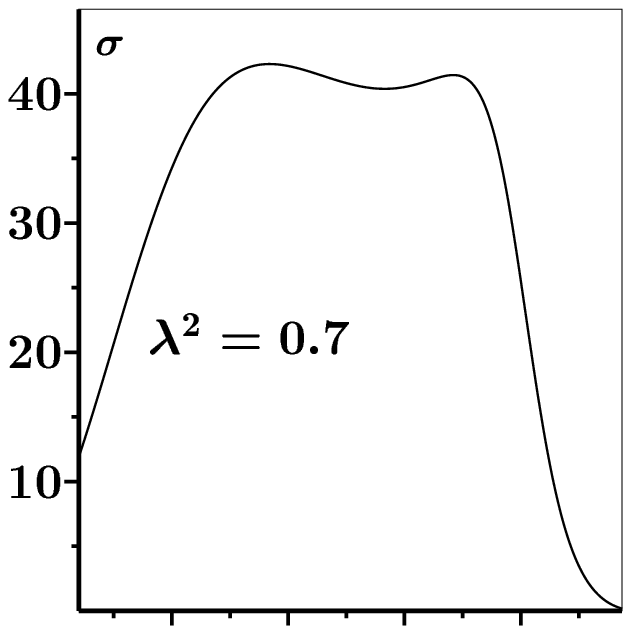}}
\scalebox{0.6}{\includegraphics
{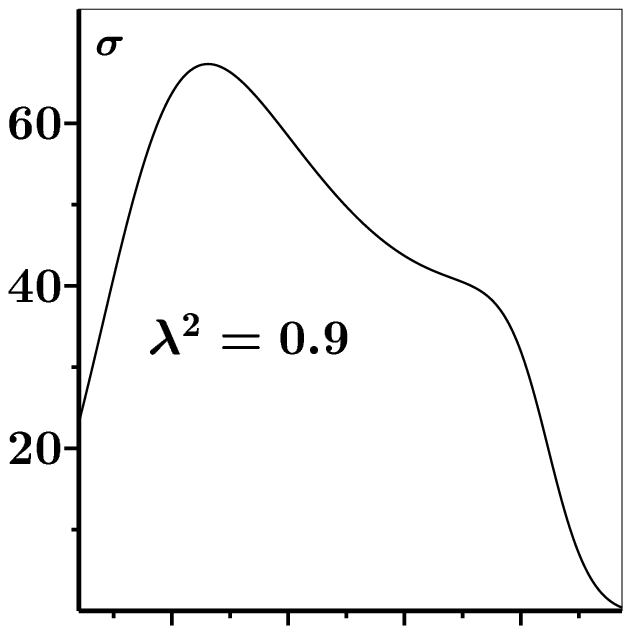}}
\scalebox{0.6}{\includegraphics
{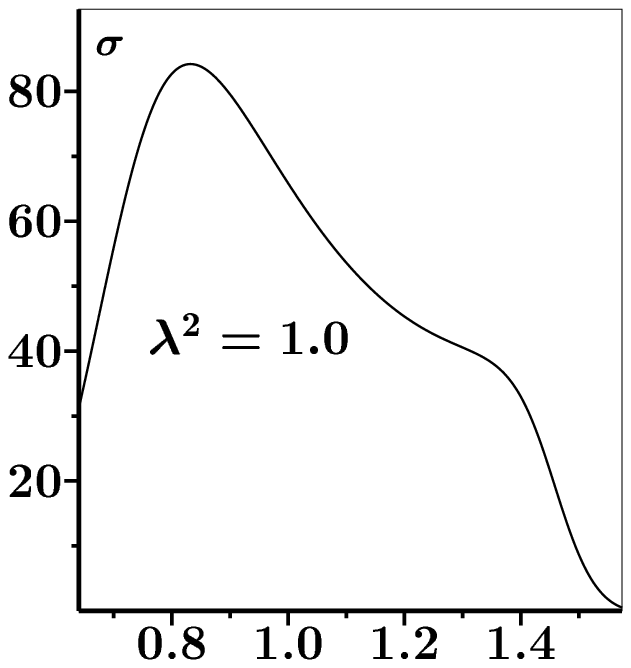}}}
\end{tabular}
\end{center}
\normalsize
\caption{Theoretical cross sections $\sigma$ in units of GeV$^{-2}$,
as a function of the strength
of the transition potential, characterised by $\lambda$,
for $K\pi$ elastic $S$-wave scattering, coupled to a confinement spectrum
with ground state at 1.31 GeV
and level spacing of 380 MeV
(model parameters taken from Ref.~\cite{EPJC22p493}).
The horizontal axes, in units of GeV, are identical for all figures,
and run from the $K\pi$ threshold up to about 1.5 GeV.}
\label{kappa_cs}
\end{figure}

In Fig.~\ref{kappa_cs} we show,
for a variety of $\lambda$ values,
the theoretical cross sections as they follow from formula (\ref{T}),
for $K\pi$ elastic $S$-wave scattering.
The ground state of the $n\bar{s}$ confinement spectrum is chosen at
1.31 GeV, a typical value in standard quark models,
whereas the level spacing equals 0.38 GeV.
Actually, we truncate the summations in formula (\ref{T})
after the second term, and parametrise the sum of the remaining terms
by a constant (see Ref.~\cite{EPJC22p493} for details).
For small coupling ($\lambda^{2}=0.1$), we obtain Breit-Wigner-shaped
resonances at about the energies of the confinement spectrum,
similar to the case of atomic spectra.
Here, we only show the ground state.
But then, as $\lambda$ increases, one observes the {\it birth}
\/of a second, lower-lying structure, which dominates at the
final flavour-independent physical value for the coupling, namely
$\lambda=1$.
The theoretical curve for $\lambda=1$ is identical to the
curve shown in Fig.~\ref{KpiS}.

We thus obtain, for free, the $K^{\ast}_{0}(800)$ structure,
just from the coupling of a $q\bar{q}$ confinement spectrum to
meson-meson scattering. This structure is clearly not related,
at least not in a simple way, to the confinement ground state,
since the $K^{\ast}_{0}(1430)$ resonance associated with the latter state
is visually well recognisable for different values of the coupling.
Consequently, the $K^{\ast}_{0}(800)$ must have a different origin.

\begin{figure}[htbp]
\begin{center}
\begin{tabular}{cc}
\resizebox{6.0cm}{!}{\includegraphics
{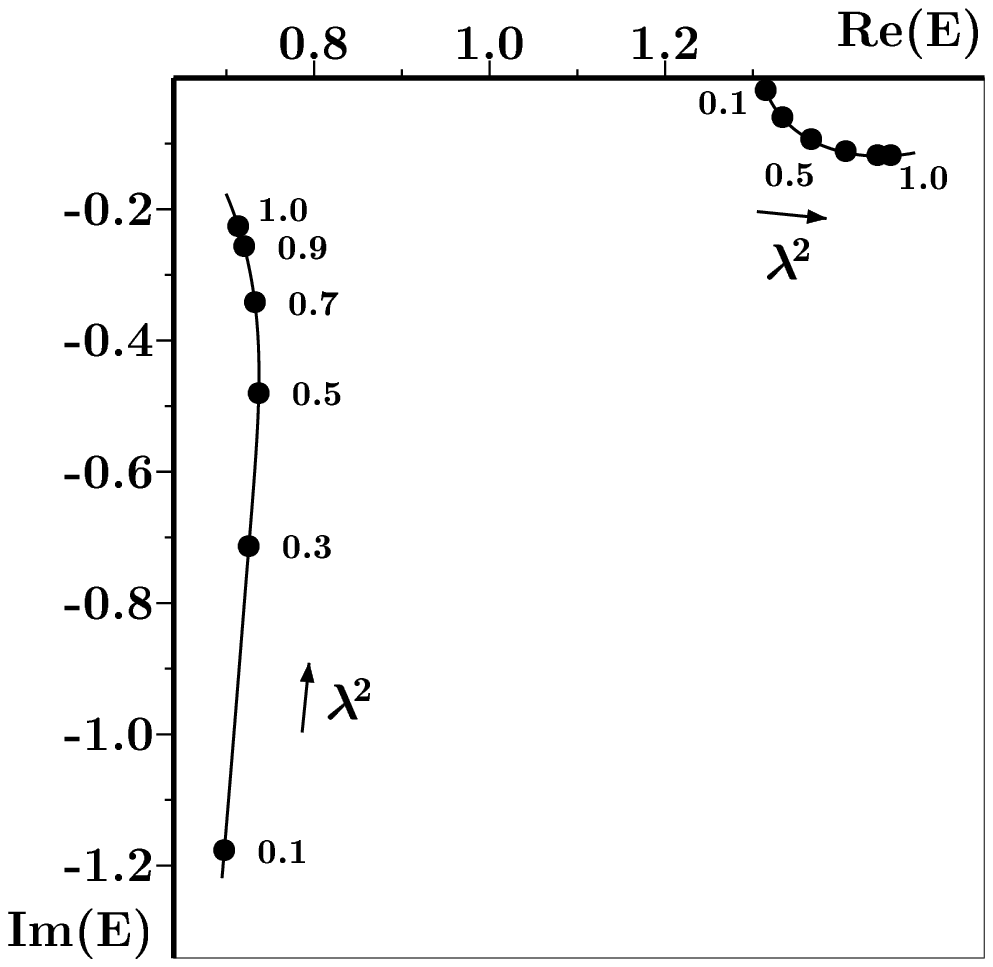}} &
\resizebox{5.0cm}{!}{\includegraphics
{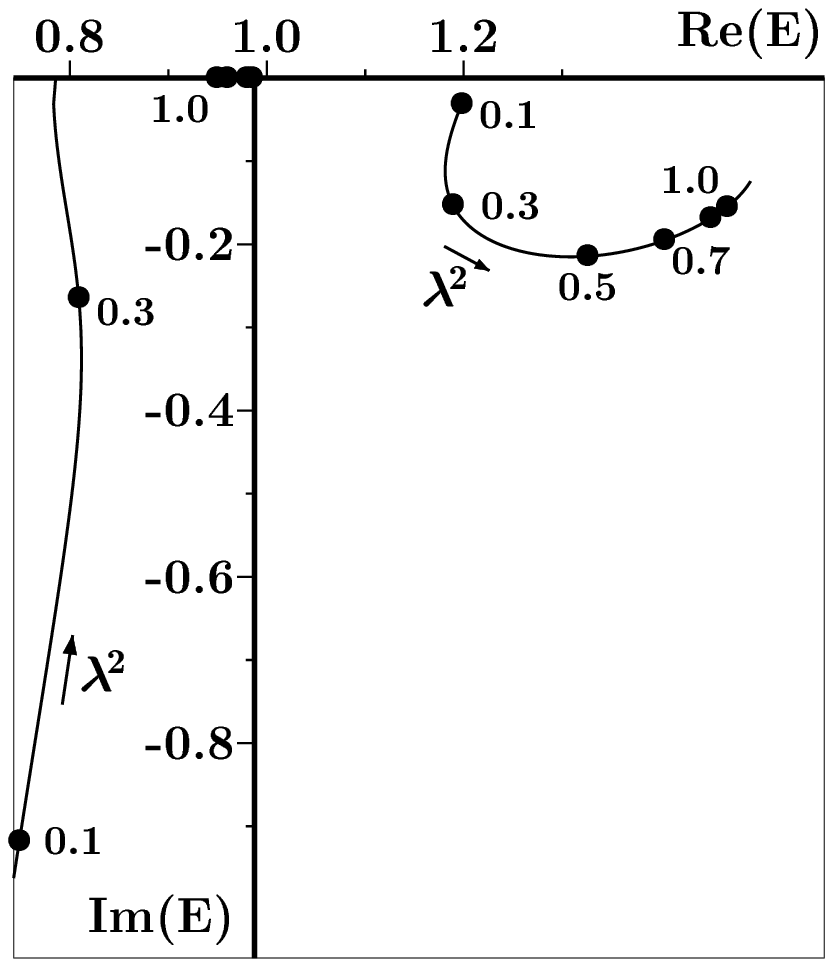}}\\
(a) & (b)
\end{tabular}
\end{center}
\caption{The scattering-matrix poles for elastic $K\pi$ (a) and $KK$ (b)
scattering in $S$ waves, as a function of the coupling constant $\lambda$.
Along the real and imaginary axes of the $E=\sqrt{s}$ complex plane,
units are in GeV.}
\label{S_poles}
\end{figure}

The information contained in Fig.~\ref{kappa_cs} can be compactified
by studying the singularities of the scattering matrix that follow from
formula (\ref{T}).
In Fig.~\ref{S_poles}a we depict, by solid circles,
those scattering-matrix pole positions in the complex-$\sqrt{s}$ plane
which dominate the shape of each of the graphs shown in Fig.~\ref{kappa_cs}.
To each graph correspond two nearby poles.
Besides these most relevant poles, there exists an infinity of poles
in the whole complex-$\sqrt{s}$ plane, for each value of the coupling
constant $\lambda$.
When $\lambda$ is varied, the pole positions draw trajectories.

In Fig.~\ref{S_poles}a one observes two distinct trajectories.
For small coupling ($\lambda^{2}=0.1$), we have the following poles:
the $K^{\ast}_{0}(800)$ pole,
with a real part close to threshold, and a large imaginary part
($\approx -1.2i$ GeV), plus
the $K^{\ast}_{0}(1430)$ pole,
with a small imaginary part,
close to the confinement ground state at 1.31 GeV.
The $K^{\ast}_{0}(1430)$ pole shows up as the sharp peak
in Fig.~\ref{kappa_cs} (upper left),
whereas the $K^{\ast}_{0}(800)$ pole
produces no significant signal in the same figure.
Actually, in the limit $\lambda\rightarrow 0$,
the $K^{\ast}_{0}(800)$ pole
disappears to negative imaginary infinity, which is
the scattering continuum, leaving no signal at all
in the theoretical $K\pi$ elastic $S$-wave scattering cross section,
whereas the $K^{\ast}_{0}(1430)$ pole approaches the real axis
at the energy of the confinement ground state, resulting in an
infinitely narrow resonance at precisely 1.31 GeV.

For increasing coupling, the $K^{\ast}_{0}(800)$ pole approaches the real
axis, resulting in a significant signal in
the $K\pi$ elastic $S$-wave scattering cross section,
whereas the $K^{\ast}_{0}(1430)$ pole moves further away from the
real axis, giving rise to a broader resonance at higher energies
(see Fig.~\ref{kappa_cs}).
For the physical coupling $\lambda=1$, we find the $K^{\ast}_{0}(1430)$ pole
at $(1.42-0.11i)$ GeV, as well as the $K^{\ast}_{0}(800)$ enhancement.

Upon substituting the constituent strange quark mass
by a nonstrange mass, which differ about 100 MeV in our model,
one obtains $E=1.21$ GeV for
the ground state of the confinement spectrum
of $n\bar{n}$ \/ isotriplet scalar mesons.
The level spacings are flavour independent, hence also 0.38 GeV in this case.
Isotriplet scalar mesons couple strongly to $K\bar{K}$.
The poles following from the scattering amplitude~(\ref{T})
for $K\bar{K}$ elastic $S$-wave scattering
are depicted in Fig.~\ref{S_poles}b.
Here we observe
a situation which is very similar to the previous case.
Only now, for the physical coupling,
the $a_{0}(980)$ pole ends up as a bound state, on the real axis,
some 20 MeV below threshold,
whereas the $a_{0}(1450)$ pole is found at $(1.46-0.16i)$ GeV.
Both poles change their locations when also $\eta\pi$ is coupled,
thus describing experiment in a realistic fashion.

Before we continue, we shall spend some words on the motion of
the $a_{0}(980)$ pole as a function of $\lambda$.
Like in the case of the $K^{\ast}_{0}(800)$ pole,
the $a_{0}(980)$ pole position has a large negative imaginary part
for small coupling, which decreases when the coupling increases.
For $\lambda^{2}\approx 0.35$, the $a_{0}(980)$ pole settles on the
real axis below the $K\bar{K}$ threshold.
In this situation, it represents a virtual bound state.
For increasing coupling, it moves towards threshold,
where it arrives for $\lambda\approx 0.5$.
Then it becomes a real bound-state pole, which starts moving
in the opposite direction, away from threshold.
For $\lambda=1.0$, it is found some 20 MeV below threshold.
The $K^{\ast}_{0}(800)$ pole moves in a similar fashion for $\lambda$ values
larger than 1.

As a final remark on Fig.~\ref{S_poles}, we draw the attention
of the reader to the very nonperturbative behaviour
of the pole trajectories.
A perturbative behaviour would be linear in $\lambda^{2}$ for the Born term,
or quadratic in this variable for the next term,
which is {\em not} \/what one observes for the pole trajectories.
So the $K^{\ast}_{0}(800)$ and $a_{0}(980)$ poles {\em cannot} \/be obtained
perturbatively.

The existence of poles in the scattering amplitude
that do not stem from the confinement spectrum
is a very important finding of the unitarised model for meson-meson scattering
\cite{ZPC30p615,PRL76p1575}.
Such poles we shall henceforth call ``extra'' poles or ``continuum'' poles.
However, a word of caution is due at this point. Namely, the coupling $\lambda$
is not the only physical parameter of the coupled-channel problem. If we vary
any of the others, certainly the pole positions will change as well. Now, it
turns out that, by choosing different yet physically very acceptable values
for some of the other parameters, the pole {\em trajectories} \/as a function
of $\lambda$ may change dramatically in situations where two poles apporach
each other sufficiently. In particular, a cross-over may take place in such
situations, implying that one pole takes over the role of the other, and
vice versa. So what appears to be a ``confinement'' pole for one set of
parameters could also be a ``continuum'' pole for another set.
Therefore, whatever name we
attribute to a certain pole, one should always keep in mind that this label
may depend on the specific choice of parameters. Nevertheless, it also turns
out that the {\em terminal points} \/of the pole trajectories, i.e., for a
fixed value of the coupling $\lambda$, are rather {insensitive} \/to moderate
changes in the other parameters, thus keeping the {\em physically observable}
\/predictions quite stable.

In Fig.~\ref{S_poles}, we show that the physics behind
the $K^{\ast}_{0}(800)$ and $a_{0}(980)$ structures
is identical.
Moreover, we find for isoscalar $J^{P}\!=\!0^{+}$ meson-meson scattering
two poles, associated with the $f_{0}(1370)$ and $f_{0}(1500)$ resonances,
which stem from the $n\bar{n}$
($n\bar{n}$ stands here for $(u\bar{u}+d\bar{d})/\sqrt{2}$)
and $s\bar{s}$ confinement spectrum, respectively.
However, we also find two lower-lying poles,
of identical origin as those described in Fig.~\ref{S_poles},
and which correspond to the $f_{0}(600)$
and $f_{0}(980)$ resonances \cite{ZPC30p615}.

Through the closed analytic form of the scattering amplitude (\ref{T}),
it is easy to classify the scalar mesons by studying the pole structure
of the resulting scattering matrix.
One nonet, viz.\ the
$f_{0}(1370)$, $a_ {0}(1450)$, $K^{\ast}_{0}(1430)$, and $f_{0}(1500)$,
stems from the $q\bar{q}$ confinement ground states,
whereas the other nonet, consisting of the
$f_{0}(600)$, $a_ {0}(980)$, $K^{\ast}_{0}(800)$, and $f_{0}(980)$,
stems from the scattering continuum.

As to the question of what these scalar mesons are composed of,
the model behind expression (\ref{T}) can only respond partly.
When we refer to quarks, we have constituent quarks in mind,
which incorporate a large fraction of the QCD interactions through
their mass parameter.
This implies that our $q\bar{q}$ systems, at a more fundamental level,
consist of valence quarks, glue, and virtual quark-antiquark pairs.
Any configuration of quarks, antiquarks, and gluons
that cannot be decomposed into a pair of colour-singlet mesons
contributes, in our model, to confinement.
Thus, mesonic systems are mixtures of constituent $q\bar{q}$ and two-meson
states in our model.
Consequently, we find that even true mesonic bound states,
which lie below all possible (OZI-allowed) two-meson thresholds,
contain virtual meson-meson pairs \cite{ZPC19p275}.
This means, for example, that the $J/\psi(3095)$ contains fractions
of virtual pairs of $D$, $D_{s}$, $D^{\ast}$, and $D^{\ast}_{s}$
mesons \cite{PRD21p772}.

How realistically this model describes Nature became evident when we determined
our prediction for the new $D^{\ast}_{sJ}(2317)$ resonance \cite{PRL91p012003},
just after it was reported by the BaBar collaboration \cite{PRL90p242001}.
Assuming a $J^{P}\!=\!0^{+}$ assignment, as suggested by BaBar,
we found poles in the $DK$ scattering amplitude of our model
very similar to the poles for $S$-wave $K\pi$ and $KK$ scattering.
In the isotriplet $J^{P}\!=\!0^{+}$ $KK$ scattering amplitude, we found
a bound-state pole on the real axis, just below the $KK$ threshold
\cite{HEPPH0304105}. Due to a small coupling to $\eta\pi$, this pole manifests
itself as a relatively narrow resonance in $\eta\pi$ scattering
\cite{ZPC30p615}. In the $DK$ scattering amplitude,
we found a bound-state pole below the $DK$ threshold, which shows up as a
very narrow resonance in $D_{s}\pi$ scattering, due to a tiny coupling to
this OZI-forbidden and isospin-violating \cite{PRD68p054006} channel.
In the isotriplet $J^{P}\!=\!0^{+}$ $KK$ scattering amplitude, we also found
a pole describing the $a_ {0}(1450)$ resonance.
Likewise, in $DK$ scattering we found such a pole describing a resonance
at about 2.8 GeV, with a width of some 400 MeV \cite{HEPPH0312078}.

Here, we now also study other flavour configurations for scalar mesons.
The principal results are collected in Table~\ref{scalars}.
\begin{table}[hbpt]
\begin{center}
\begin{tabular}{|c|c||c||c|c|}
\hline & & & & \\ [-8pt]
& & {\it ``continuum''} state & {\it ``ground''} state & higher recurrence\\
channel & $q\bar{q}$ & $\sqrt{s}$ (GeV) & $\sqrt{s}$ (GeV) & $\sqrt{s}$ (GeV)\\
\hline & & & & \\ [-8pt]
$\pi\pi$ & $n\bar{n}$ & $0.47-0.21i$ & $1.36-0.13i$ & -\\
\hline & & & & \\ [-8pt]
$K\pi$ & $s\bar{n}$ & $0.73-0.26i$ & $1.46-0.12i$ & $1.71-0.02i$\\
\hline & & & & \\ [-8pt]
$\eta\pi$ & $n\bar{n}$ & $0.97-0.028i$ & $1.45-0.13i$ & -\\
\hline & & & & \\ [-8pt]
$\pi\pi$ & $s\bar{s}$ & $0.99-0.020i$ & $1.51-0.06i$ & -\\
\hline & & & & \\ [-8pt]
$D\pi$ & $\bar{c}u/d$ & $2.14-0.16i$ & $2.58-0.12i$ & -\\
\hline & & & & \\ [-8pt]
$DK$ & $c\bar{s}$ & 2.33 & $2.80-0.20i$ & -\\
\hline & & & & \\ [-8pt]
$B\pi$ & $\bar{b}u/d$ & $6.06-0.29i$ & $5.46-0.03i$ & $6.03-0.05i$\\
\hline & & & & \\ [-8pt]
$BK$ & $b\bar{s}$ & $6.21-0.33i$ & 5.61 & $6.05-0.03i$\\
\hline & & & & \\ [-8pt]
$BD$ & $\bar{b}c$ & $7.12-0.43i$ & 6.64 & $7.11-0.03i$\\
\hline
\end{tabular}
\end{center}
\caption{Bound states and resonance poles for meson-meson elastic
$S$-wave scattering
\cite{ZPC30p615,EPJC22p493,HEPPH0304105,HEPPH0312078,InPreparation}.}
\label{scalars}
\end{table}
At this point, one should realise that, when studying heavier mesons,
especially
those with bottom quarks, flavour independence of the strong interactions
\cite{PRD59p012002} does not only have implications for the level spacings
of the
confinement spectrum, but also for the parameters $a$ and $\lambda$ of the
transition mechanism. Thus, we take \cite{HEPPH0310320}
\begin{equation}
a_{xy}\,\sqrt{\mu_{xy}}\; =\;\xrm{constant}
\;\;\;\;\xrm{and}\;\;\;\;
\lambda_{xy}\,\sqrt{\mu_{xy}}\; =\;\xrm{constant}
\;\;\;\; ,
\label{flavorinvariance}
\end{equation}
where $x$ and $y$ represent the two flavours involved,
and $\mu$ the reduced quark mass.
The constants in Eq.~(\ref{flavorinvariance}) are fixed by
\cite{HEPPH0110156} $a_{us}=3.2$ GeV$^{-1}$ and
$\lambda_{us}=0.75$ GeV$^{-3/2}$ (throughout this work set to unity),
which parameters have been adjusted to $S$-wave $K\pi$ scattering.
In Ref.~\cite{PRL91p012003} this scaling had not been taken into account.
As a consequence, the pole positions for $D\pi$ and $DK$ in the present work
differ slightly from our previous results.
In particular, the $D_{s0}^{\ast}$ pole position now comes out
only 10 MeV above the experimental mass of 2.317 GeV.

\begin{figure}[htbp]
\begin{center}
\begin{tabular}{cc}
\resizebox{5.0cm}{!}{\includegraphics
{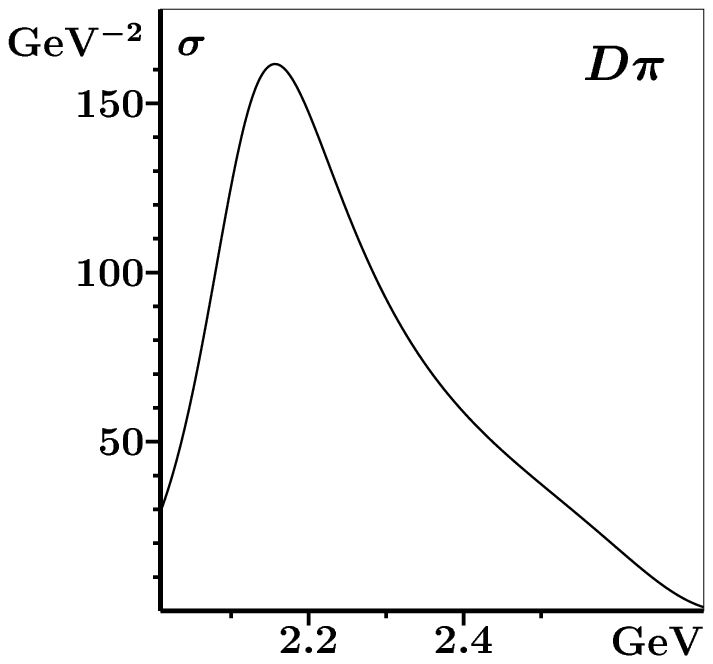}} &
\resizebox{5.0cm}{!}{\includegraphics
{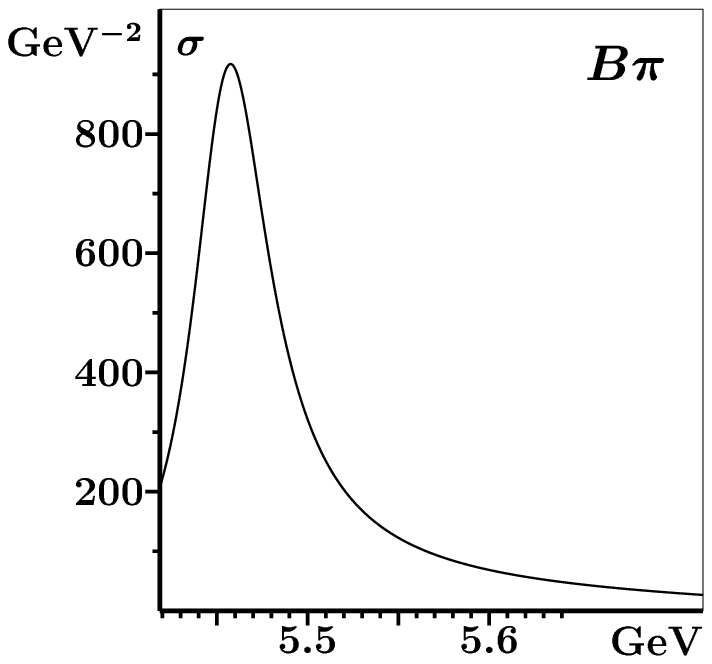}}\\
(a) & (b)
\end{tabular}
\end{center}
\caption{The theoretical scattering cross section for elastic
$S$-wave $D\pi$ (a) and $B\pi$ (b) scattering near threshold.}
\label{DBncross}
\end{figure}

For the lowest-lying structures in the $D\pi$ and $B\pi$ systems,
we find resonances close to threshold, as depicted in Fig.~\ref{DBncross}.
So, within the accuracy of the model, we may predict a broad resonance
in $D\pi$, peaking at 2.16$\pm$0.05 GeV and with a width of 300$\pm$50 MeV,
possibly seen in experiment \cite{HEPEX0307021,HEPEX0210037},
whereas in $B\pi$ a narrow resonance has to be expected
at 5.47$\pm$0.05 GeV, with a width of 50$\pm$20 MeV.
The latter two resonances have very different scattering widths,
because they are relate to different phenomena:
the $D_{0}^{\ast}$(2160) resonance originates from
the scattering continuum, whereas the foreseen $B_{0}^{\ast}$(5470) stems from
the ground state of the $J^{P}=0^{+}$ $\bar{b}n$ confinement spectrum.
In fact, for mesons with $b$ quarks, the lowest-lying ``continuum'' poles
come out with real parts larger than those of the first radial excitations,
and moreover with a large imaginary part.
The effect of these poles will not easily be distinguished from the
scattering background.

The lowest-lying scalar states $B_{s0}^{\ast}$(5610) in $BK$,
and $B_{c0}^{\ast}$(6640) in $BD$, both come out well below the
respective OZI-allowed thresholds, and are thus considered pure bound states
in our model, though having sufficient phase space for the OZI-forbidden
decays $B_{s}\pi$ and $B_{c}\pi$ (for $M_{B_{c}}\approx$ 6.4 GeV),
respectively, similarly to the $D_{s0}^{\ast}(2317)$ state in $DK$.

\section{The classification of meson poles}

We may conclude that, in the non-exotic sector of meson-meson scattering,
resonances are well described by a coupled-channel model
which incorporates OZI-allowed transitions to
the $q\bar{q}$ confinement sector.
The model, moreover, yields bound states for masses
below the lowest possible meson-meson decay channel,
which can be measured via alternative scattering or decay processes.

The analytic continuation to complex $\sqrt{s}$
of the resulting scattering amplitude contains poles.
These can be grouped into flavour multiplets,
without addressing the question whether any observable resonances are
associated with them.
While the same pole may appear as a sharp resonance
for one flavour configuration, it could give rise to a broad structure,
possibly not even a resonance, for another.
Within the model, one may verify whether poles are of the same origin
by continuously varying the involved mass parameters from
one flavour to another.

We observed that the poles describing the
$f_{0}(600)$ and $K^{\ast}_{0}(800)$ enhancements,
the $a_ {0}(980)$ and $f_{0}(980)$ resonances,
and the $D^{\ast}_{s0}(2317)$ bound state
(bound with respect to OZI-allowed decays)
are the lowest-lying poles in the scattering amplitudes
for the respective flavour and isospin configurations,
which, furthermore, all disappear into the scattering continuum
for vanishing coupling $\lambda$.
Therefore, we conclude that all these states have the same origin,
and constitute as such one big flavour multiplet.

For the scalar-meson resonances even higher in mass,
several similar poles are to be expected,
thus almost doubling the number of resonances
with respect to the confinement spectrum.
Modelling the corresponding scattering amplitudes is complicated,
not for some fundamental difficulty,
but just because of the large number of meson-meson channels involved,
with very disparate thresholds, some containing highly unstable mesons in
their turn.
Nevertheless, the model as it stands is very much capable of
indicating the global classification of mesonic resonances.
But it shows that things are just becoming slightly more intricate than
when Balmer did the job for atomic spectra.
Of course, many other details remain to be addressed.

The discovery of the $D^{\ast}_{s0}(2317)$ meson
\cite{PRL90p242001,HEPEX0305100,HEPEX0308019}
turned out to be of great significance in the process of classifying
the $f_{0}(600)$ and $K^{\ast}_{0}(800)$ enhancements
within the same flavour multiplet as
the $a_ {0}(980)$ and $f_{0}(980)$ resonances.

\section*{Acknowledgments}

This work was partly supported by the
{\it Funda\c{c}\~{a}o para a Ci\^{e}ncia e a Tecnologia}
of the {\it Minist\'{e}rio da
Ci\^{e}ncia e do Ensino Superior} \/of Portugal,
under contract numbers
POCTI/\-35304/\-FIS/\-2000
and
POCTI/\-FNU/\-49555/\-2002.

\end{document}